\def\ra{\rangle}
\def\la{\langle}

\def\be{\begin{equation}}
\def\ee{\end{equation}}
\def\ba{\begin{array}}
\def\ea{\end{array}}

\def\ra{\rangle}
\def\la{\langle}

\documentstyle[11pt]{article}
\input amssym.def
\begin{document}
\baselineskip=18pt \setcounter{page}{1}
\begin{center}
{\LARGE \bf Canonical Form and  Separability of PPT States in ${\cal C}^2 \otimes
{\cal C}^2\otimes {\cal C}^2 \otimes {\cal C}^N$ Composite Quantum
Systems}
\end{center}
\vskip 2mm

\begin{center}
{\normalsize Shao-Ming Fei$^{1,\  2}$, Xiu-Hong Gao$^1$, Xiao-Hong
Wang$^1$, Zhi-Xi Wang$^1$, and Ke Wu$^1$  }
\medskip

\begin{minipage}{4.8in}
{\small \sl $ ^1$
Department of Mathematics, Capital  Normal University, Beijing,
China.}\\
{\small \sl $ ^2$ Institute of Applied Mathematics,
University of Bonn,  53115 Bonn Germany}
\end{minipage}
\end{center}

\vskip 2mm
\begin{center}

\begin{minipage}{4.8in}

\centerline{\bf Abstract}
\bigskip

We give a canonical form of PPT states in ${\cal C}^2
\otimes {\cal C}^2\otimes {\cal C}^2 \otimes {\cal C}^N$ with
rank=$N$. From this canonical form a necessary separable condition
for these states is presented.

\vskip 9mm
Key words: Separability, Quantum entanglement
\vskip 1mm
PACS number(s): 03.67.Hk, 03.65.Ta, 89.70.+c
\end{minipage}
\end{center}

\vskip0.4cm

Quantum entangled states have become one of the key resources in
the rapidly expanding field of quantum information processing and
computation \cite{DiVincenzo,teleport,dense,crypto}. Nevertheless,
the physical character and mathematical structure of the quantum
entanglement are still not fully understood yet. One even does not
know wether a general quantum (mixed) state is entangled or not,
and how entangled it is after some noisy quantum processes.

To quantify entanglement, a number of entanglement measures such
as entanglement of formation and distillation
\cite{568,7}, negativity
\cite{Peres96a}, von Neumann entropy and relative
entropy \cite{568,sw} have been proposed for bipartite
states. However most proposed measures of entanglement involve
extremizations which are difficult to handle analytically. For
instance, the entanglement of formation \cite{17} is intended to
quantify the amount of quantum communication required to create a
given bipartite state. So far no explicit analytic formulae for
entanglement of formation have been found for systems larger than
a pair of qubits \cite{HillWootters}, except for some symmetric
states \cite{th} and a class of special states \cite{ent3}.

The separability problem concerns both bipartite and
multipartite quantum states, although the measure of entanglement
is only well defined for bipartite case. For pure states the
separability is quite well understood \cite{peresbook}.
Nevertheless, in real conditions, due to the interactions with
environment, one encounters mixed states rather than pure ones.
These mixed states can still possess some residual entanglement,
but the quantum correlations are weakened. Hence the
manifestations of mixed-state entanglement can be very subtle
\cite{10}.

A state of a composite quantum system is said to be disentangled
or separable if it can be prepared in a local or classical way.
Let ${\cal C}^M_{\sc a}$ (resp. ${\cal C}^N_{\sc b}$)
be $M$ (resp. $N$) dimensional
complex Hilbert spaces associated with sub-quantum systems ${\sc
a}$ (resp. ${\sc b}$). A separable bipartite state in
${\cal C}^M_{\sc a} \otimes {\cal C}^N_{\sc b}$ can be prepared as an
ensemble realization of pure product states $\left\vert
\psi^{i}_{\sc a}\right\rangle \left\vert \psi^{i}_{\sc
b}\right\rangle$ occurring with a certain probability $p_{i}$:
\begin{equation}
\rho _{\sc ab}=\sum_{i}p_{i}\rho^{i}_{\sc a}\otimes \rho^{i}_{\sc b},
\label{sep2}
\end{equation}
where $\sum_{i}p_{i}=1$,
$\rho^{i}_{\alpha}=\left\vert \psi^{i}_{\alpha}\right\rangle
\left\langle \psi^{i}_{\alpha}\right\vert$,
$\left\vert \psi^{i}_\alpha\right\rangle$
are normalized pure
states associated with the subsystems $\alpha$,
$\left\langle \psi^{i}_{\alpha}\right\vert$ are the transpose and conjugate of
$\left\vert \psi^{i}_{\alpha}\right\rangle$,
$\alpha={\sc a,b}$. If no convex
linear combination exists for a given $\rho _{\sc ab}$, the state is
called entangled.

For a generic mixed state $\rho _{\sc ab}$, finding a
decomposition like in Eq.~(\ref{sep}) or proving that it does not
exist is a non-trivial task (see\cite{lbck00}
and references therein). The Bell inequalities satisfied by a
separable system give the first necessary condition for
separability \cite{Bell64}. Afterwards the Peres criterion
\cite{Peres96a} says that partial transpositions with respect to
one or more subsystems of a separable state $\rho$ are positive:
$\rho ^{t_{\alpha}}\geq 0$, where $\alpha$ is either ${\sc
a}$ or ${\sc b}$, $t_{\alpha}$ stands for the partial
transposition with respect to a subsystem $\alpha$. This criterion
was further shown to be also sufficient for bipartite systems in
${\cal C}^2\otimes {\cal C}^2$ and ${\cal C}^2\otimes {\cal C}^3$
\cite{3hPLA223}. The reduction criterion proposed independently in
\cite{2hPRA99} and \cite{cag99} gives another necessary criterion
which is equivalent to the Peres criterion for ${\cal C}^2\otimes
{\cal C}^N$ composite systems but is generally weaker. There are
many other necessary criteria such as majorization
\cite{nielson01}, entanglement witnesses
\cite{3hPLA223,ter00}, extension of Peres criterion \cite{dps}, matrix
realignment \cite{ru02}, generalized partial transposition
criterion (GPT) \cite{chenPLA02}, generalized reduced criterion
\cite{grc}. For low rank density matrices there are also some
necessary and sufficient criteria of separability
\cite{hlpre}.

The separability and entanglement in ${\cal C}^2\otimes {\cal
C}^2\otimes {\cal C}^N$ and ${\cal C}^2\otimes {\cal C}^3\otimes
{\cal C}^N$ composite quantum systems have been studied in terms
of matrix analysis on tensor spaces \cite{22n}. It is shown
that all such quantum states $\rho$ with positive partial
transposes and rank $r(\rho)\leq N$ are separable. In this article
we extend the results in \cite{22n} to the case of composite
quantum systems in ${\cal C}^2 \otimes {\cal C}^2\otimes {\cal
C}^2 \otimes {\cal C}^N$. We give a canonical form of positive
partial transpose (PPT) states in
${\cal C}^2 \otimes {\cal C}^2\otimes {\cal C}^2 \otimes {\cal
C}^N$ with rank $N$ and present a necessary separability criterion.

A separable state in
${\cal C}^2_{\sc a} \otimes {\cal C}^2_{\sc b} \otimes {\cal
C}^2_{\sc c} \otimes {\cal C}^N_{\sc d}$ is of the form:
\begin{equation}
\rho _{\sc abcd}=\sum_{i}p_{i}\rho^{i}_{\sc a}\otimes \rho^{i}_{\sc b}
\otimes \rho^{i}_{\sc c}\otimes \rho^{i}_{\sc d},
\label{sep}
\end{equation}
where $\sum_{i}p_{i}=1$, $0< p_i\leq 1$,
$\rho^{i}_{\alpha}$ are desity matrices associated with the subsystems $\alpha$,
$\alpha={\sc a,b,c,d}$.
In the following we denote by $R(\rho)$, $K(\rho)$, $r(\rho)$ and
$k(\rho)$ the range, kernel, rank, dimension of the
kernel of $\rho$, respectively.

We first derive a canonical form of PPT states in
${\cal C}^2_{\sc a} \otimes {\cal C}^2_{\sc b}\otimes {\cal C}^2_{\sc c} \otimes {\cal
C}^N_{\sc d}$ with rank $N$, which allows for an explicit
decomposition of a given state in terms of convex sum of
projectors on product vectors. Let
$|0_A\ra$, $|1_A\ra$; $|0_B\ra$, $|1_B\ra$;
$|0_C\ra$, $|1_C\ra$ and $|0_D\ra\, \cdots \,|N-1_D\ra$ be some local bases
of the sub-systems ${\sc a,~b,~c,~d}$ respectively.

{\bf Lemma. }\ \  Every PPT state $\rho$ in ${\cal C}^2_{\sc a}
\otimes {\cal C}^2_{\sc b}\otimes {\cal C}^2_{\sc c} \otimes {\cal
C}^N_{\sc d}$ such that
$r(\la 1_A, 1_B,1_C|\rho |1_A, 1_B,1_C\ra)=r(\rho)=N$, can be
transformed into the following
canonical form by using a reversible local operation:
\be
\rho=\sqrt{D}[CBA\ \ CB\ \ CA\ \ C \ \ BA\ \ B\
\ A\ \ I]^{\dag} [CBA\ \ CB\ \ CA\ \ C \ \ BA\ \ B\ \ A\ \
I]\sqrt{D}
\ee
where $A$, $B$, $C$, $D$ and the identity $I$ are $N\times N$ matrices
acting on ${\cal C}_{\sc d}^N$ and satisfy the following relations:
$[A,\ A^{\dag}]=[B,\ B^{\dag}]=[C,\
C^{\dag}]=[B,\ A]=[B,\ A^{\dag}] =[C,\ A]=[C,\ A^{\dag}]=[C,\
B]=[C,\ B^{\dag}]=0$ and $D=D^{\dag}$ ($\dag$ stands for the transpose
and conjugate).

{\bf Proof.}\ In the considered
basis a density matix $\rho$ can be always written as:
\be\label{r}
\rho=\left(
\begin{array}{cccccccc}
E_1&E_{12}&E_{13}&E_{14}&E_{15}&E_{16}&E_{17}&E_{18}\\
E_{12}^{\dag}&E_2&E_{23}&E_{24}&E_{25}&E_{26}&E_{27}&E_{28}\\
E_{13}^{\dag}&E_{23}^{\dag}&E_{3}&E_{34}&E_{35}&E_{36}&E_{37}&E_{38}\\
E_{14}^{\dag}&E_{24}^{\dag}&E_{34}^{\dag}&E_4&E_{45}&E_{46}&E_{47}&E_{48}\\
E_{15}^{\dag}&E_{25}^{\dag}&E_{35}^{\dag}&E_{45}^{\dag}&E_5&E_{56}&E_{57}&E_{58}\\
E_{16}^{\dag}&E_{26}^{\dag}&E_{36}^{\dag}&E_{46}^{\dag}&E_{56}^{\dag}&E_6&E_{67}&E_{68}\\
E_{17}^{\dag}&E_{27}^{\dag}&E_{37}^{\dag}&E_{47}^{\dag}&E_{57}^{\dag}&E_{67}^{\dag}&E_7&E_{78}\\
E_{18}^{\dag}&E_{28}^{\dag}&E_{38}^{\dag}&E_{48}^{\dag}&E_{58}^{\dag}&E_{68}^{\dag}&E_{78}^{\dag}&E_8\\
\end{array}
\right),
\ee
where $E's$ are $N \times N$ matrices, $r(E_8)=N$.
After the projection $\tilde{\rho}=\la 1_A|\rho|1_A\ra$,
 we obtain
\be
\tilde{\rho}=\la1_A|\rho|1_A\ra
=\left(
\begin{array}{cccc}
E_5&E_{56}&E_{57}&E_{58}\\
E_{56}^{\dag}&E_6&E_{67}&E_{68}\\
E_{57}^{\dag}&E_{67}^{\dag}&E_7&E_{78}\\
E_{58}^{\dag}&E_{68}^{\dag}&E_{78}^{\dag}&E_8
\end{array}
\right).
\ee

$\tilde{\rho}$ is now a
state in ${\cal C}^2_{\sc b} \otimes{\cal C}^2_{\sc c} \otimes {\cal C}^N_{\sc d}$ with
$r(\tilde{\rho})=r(\rho)=N$. As
every principal minor determinant of
$\tilde \rho^{t_B}$ ($\tilde \rho^{t_C}$) is some principal minor
determinant of $\rho$, the fact that $\rho$ is PPT implies that $\tilde \rho$ is also
PPT, $\tilde{\rho}\ge 0$. Using the Lemma 1 in \cite{22n} we have
\be
\tilde{\rho}=\left(
\begin{array}{cccc}
A^{\dag}B^{\dag}BA&A^{\dag}B^{\dag}B&A^{\dag}B^{\dag}A&A^{\dag}B^{\dag}\\
B^{\dag}BA&B^{\dag}B&B^{\dag}A&B^{\dag}\\
A^{\dag}BA&A^{\dag}B&A^{\dag}A&A^{\dag}\\ BA&B&A&1
\end{array}
\right),
\ee
where $[A,A^{\dag}]=[B,B^{\dag}]=[B,A]=[B,A^{\dag}]=0$. It is direct to verify
that the following vectors in ${\cal C}^2_{\sc b}
\otimes{\cal C}^2_{\sc c} \otimes {\cal C}^N_{\sc d}$ are kernel vectors
$k(\tilde \rho)$:
$$
\ba{ll}
|\psi_f\ra=|10\ra|f\ra-|11\ra A|f\ra,&~~~~|\psi_g\ra=|01\ra|g\ra-|11\ra B|g\ra,\\[2mm]
|\psi_h\ra=|00\ra|h\ra-|11\ra BA|h\ra,&~~~~\forall |f\ra,~ |g\ra,~|h\ra\in {\cal C}^N_{\sc d}.
\ea
$$
Similarly, if we consider the projection $\la
1_B|\rho|1_B\ra$ and $\la 1_C|\rho|1_C\ra$, we have
$$
\ba{rcl}
\bar{\rho}&=&\la1_B|\rho|1_B\ra\\[5mm]
&=&\left(
\begin{array}{cccc}
E_3&E_{34}&E_{37}&E_{38}\\
E_{34}^{\dag}&E_4&E_{47}&E_{48}\\
E_{37}^{\dag}&E_{47}^{\dag}&E_7&E_{78}\\
E_{38}^{\dag}&E_{48}^{\dag}&E_{78}^{\dag}&E_8
\end{array}
\right)
=\left(
\begin{array}{cccc}
A^{\dag}C^{\dag}CA&A^{\dag}C^{\dag}C&A^{\dag}C^{\dag}A&A^{\dag}C^{\dag}\\
C^{\dag}CA&C^{\dag}C&C^{\dag}A&C^{\dag}\\
A^{\dag}CA&A^{\dag}C&A^{\dag}A&A^{\dag}\\ CA&C&A&1
\end{array}
\right)
\ea
$$
and
$$
\ba{rcl}
\check{\rho}&=&\la1_C|\rho|1_C\ra\\[5mm]
&=&\left(
\begin{array}{cccc}
E_2&E_{24}&E_{26}&E_{28}\\
E_{24}^{\dag}&E_4&E_{46}&E_{48}\\
E_{26}^{\dag}&E_{46}^{\dag}&E_6&E_{68}\\
E_{28}^{\dag}&E_{48}^{\dag}&E_{68}^{\dag}&E_8
\end{array}
\right)
=\left(
\begin{array}{cccc}
B^{\dag}C^{\dag}CB&B^{\dag}C^{\dag}C&B^{\dag}C^{\dag}B&B^{\dag}C^{\dag}\\
C^{\dag}CB&C^{\dag}C&C^{\dag}B&C^{\dag}\\
B^{\dag}CB&B^{\dag}C&B^{\dag}B&B^{\dag}\\CB&C&B&1
\end{array}
\right)\,,
\ea
$$
where $[C,C^{\dag}]=[A,C]=[A,C^{\dag}]=0$,
$[B,B^{\dag}]=[B,C]=[B,C^{\dag}]=0$.

Hence the matrix (\ref{r}) now has the form:
\be\label{rm}
\rho=\left(
\begin{array}{cccccccc}
E_1&E_{12}&E_{13}&E_{14}&E_{15}&E_{16}&E_{17}&E_{18}\\
E_{12}^{\dag}&B^{\dag}C^{\dag}CB&E_{23}&B^{\dag}C^{\dag}C
&E_{25}&B^{\dag}C^{\dag}B&E_{27}&B^{\dag}C^{\dag}\\
E_{13}^{\dag}&E_{23}^{\dag}&A^{\dag}C^{\dag}CA&A^{\dag}C^{\dag}C
&E_{35}&E_{36}&A^{\dag}C^{\dag}A&A^{\dag}C^{\dag}\\
E_{14}^{\dag}&C^{\dag}CB&C^{\dag}CA&C^{\dag}C&E_{45}
&C^{\dag}B&C^{\dag}A&C^{\dag}\\
E_{15}^{\dag}&E_{25}^{\dag}&E_{35}^{\dag}&E_{45}^{\dag}&A^{\dag}B^{\dag}BA
&A^{\dag}B^{\dag}B&A^{\dag}B^{\dag}A&A^{\dag}B^{\dag}\\
E_{16}^{\dag}&B^{\dag}C^{\dag}B&E_{36}^{\dag}&B^{\dag}C&
B^{\dag}BA&B^{\dag}B&B^{\dag}A&B^{\dag}\\
E_{17}^{\dag}&E_{27}^{\dag}&A^{\dag}CA&A^{\dag}C&
A^{\dag}BA&A^{\dag}B&A^{\dag}A&A^{\dag}\\
E_{18}^{\dag}&CB&CA&C&BA&B&A&1\\
\end{array}
\right),
\ee
It has the following kernel vectors:
\be
\ba{ll}
|001\ra|f\ra-|111\ra CB|f\ra,&~~~~|010\ra|g\ra-|111\ra CA|g\ra,\\[3mm]
|011\ra|h\ra-|111\ra C|h\ra,&~~~~|100\ra|p\ra-|111\ra BA|p\ra,\\[3mm]
|101\ra|q\ra-|111\ra B|q\ra,&~~~~|110\ra|m\ra-|111\ra A|m\ra,
\ea
\ee
for all $|f\ra,~|g\ra,~\cdots,~ |m\ra\in
{\cal C}^N_{\sc d}$. This implies
\be\label{imp}
\ba{lll}
E_{27}=B^{\dag}C^{\dag}A,&~~E_{36}=A^{\dag}C^{\dag}B,&~~
E_{25}=B^{\dag}C^{\dag}BA,\\[3mm]
E_{35}=A^{\dag}C^{\dag}BA,&~~
E_{45}=C^{\dag}BA,&~~E_{23}=B^{\dag}C^{\dag}CA,\\[3mm]
E_{17}=E_{18}A,&~~ E_{16}=E_{18}B,&~~ E_{15}=E_{18}BA,\\[3mm]
E_{14}=E_{18}C,&~~ E_{13}=E_{18}CA,&~~E_{12}=E_{18}CB.
\ea
\ee

Substituting (\ref{imp}) into (\ref{rm}) and
consider partial transpose of $\rho$ with respect to the
first sub-system ${\sc a}$, we have
\be
\rho^{t_{\sc a}}=\left(
\begin{array}{cccccccc}
E_1&E_{18}CB&E_{18}CA&E_{18}C&A^{\dag}B^{\dag}E_{18}^{\dag}
&A^{\dag}B^{\dag}CB&A^{\dag}B^{\dag}CA&A^{\dag}B^{\dag}C\\
B^{\dag}C^{\dag}E_{18}^{\dag}&B^{\dag}C^{\dag}CB&B^{\dag}C^{\dag}CA&
B^{\dag}C^{\dag}C&B^{\dag}E_{18}^{\dag}&B^{\dag}CB&B^{\dag}CA&B^{\dag}C\\
A^{\dag}C^{\dag}E_{18}^{\dag}&A^{\dag}C^{\dag}CB&A^{\dag}C^{\dag}CA&
A^{\dag}C^{\dag}C&A^{\dag}E_{18}^{\dag}&A^{\dag}CB&A^{\dag}CA&A^{\dag}C\\
C^{\dag}E_{18}^{\dag}&C^{\dag}CB&C^{\dag}CA&C^{\dag}C&E_{18}^{\dag}
&CB&CA&C\\
E_{18}BA&E_{18}B&E_{18}A&E_{18}&A^{\dag}B^{\dag}BA
&A^{\dag}B^{\dag}B&A^{\dag}B^{\dag}A&A^{\dag}B^{\dag}\\
B^{\dag}C^{\dag}BA&B^{\dag}C^{\dag}B&B^{\dag}C^{\dag}A&B^{\dag}C^{\dag}
&B^{\dag}BA&B^{\dag}B&B^{\dag}A&B^{\dag}\\
A^{\dag}C^{\dag}BA&A^{\dag}C^{\dag}B&A^{\dag}C^{\dag}A&A^{\dag}C^{\dag}
&A^{\dag}BA&A^{\dag}B&A^{\dag}A&A^{\dag}\\
C^{\dag}BA&C^{\dag}B&C^{\dag}A&C^{\dag}&BA&B&A&1
\end{array}
\right).
\ee

Since the partial transpose with respect to the sub-system ${\sc a}$
is positive, $\rho^{t_{\sc a}}\ge 0$, and it does not change
$\la 1_{\sc a}|\rho|1_{\sc a}\ra$, we still have
$|100\ra|p\ra-|111\ra BA|p\ra\in k(\rho^{t_{\sc a}})$.
This gives rise to the following equalities:
$E_{18}^{\dag}=CBA$, $E_{18}=A^{\dag}B^{\dag}C^{\dag}$. $\rho$ is then of the
following form:
$$
\ba{l}
\left(
\begin{array}{cccccccc}
E_1\!&\!A^{\dag}B^{\dag}C^{\dag}CB\!&\!A^{\dag}B^{\dag}C^{\dag}CA\!&\!
A^{\dag}B^{\dag}C^{\dag}C\!&\!A^{\dag}B^{\dag}C^{\dag}BA
\!&\!A^{\dag}B^{\dag}C^{\dag}B\!&\!A^{\dag}B^{\dag}C^{\dag}A
\!&\!A^{\dag}B^{\dag}C^{\dag}\\
B^{\dag}C^{\dag}CBA\!&\!B^{\dag}C^{\dag}CB\!&\!B^{\dag}C^{\dag}CA\!&\!
B^{\dag}C^{\dag}C\!&\!B^{\dag}C^{\dag}BA\!&\!B^{\dag}C^{\dag}B\!&\!B^{\dag}C^{\dag}A
\!&\!B^{\dag}C^{\dag}\\
A^{\dag}C^{\dag}CBA\!&\!A^{\dag}C^{\dag}CB\!&\!A^{\dag}C^{\dag}CA\!&\!
A^{\dag}C^{\dag}C\!&\!A^{\dag}C^{\dag}BA\!&\!A^{\dag}C^{\dag}B\!&\!A^{\dag}C^{\dag}A
\!&\!A^{\dag}C^{\dag}\\
C^{\dag}CBA\!&\!C^{\dag}CB\!&\!C^{\dag}CA\!&\!
C^{\dag}C\!&\!C^{\dag}BA\!&\!C^{\dag}B\!&\!C^{\dag}A\!&\!C^{\dag}\\
A^{\dag}B^{\dag}CBA\!&\!A^{\dag}B^{\dag}CB\!&\!A^{\dag}B^{\dag}CA\!&\!
A^{\dag}B^{\dag}C\!&\!A^{\dag}B^{\dag}BA\!&\!A^{\dag}B^{\dag}B\!&\!A^{\dag}B^{\dag}A
\!&\!A^{\dag}B^{\dag}\\
B^{\dag}CBA\!&\!B^{\dag}CB\!&\!B^{\dag}CA\!&\!
B^{\dag}C\!&\!B^{\dag}BA\!&\!B^{\dag}B\!&\!B^{\dag}A\!&\!B^{\dag}\\
A^{\dag}CBA\!&\!A^{\dag}CB\!&\!A^{\dag}CA\!&\!
A^{\dag}C\!&\!A^{\dag}BA\!&\!A^{\dag}B\!&\!A^{\dag}A\!&\!A^{\dag}\\
CBA\!&\!CB\!&\!CA\!&\!C\!&\!BA\!&\!B\!&\!A\!&\!1
\end{array}
\right)\\[20mm]
=[\begin{array}{cccccccc}
CBA& CB&CA&C&BA&B&A&I\end{array}]^{\dag}
[CBA\,\,\, CB\,\,\, CA\,\,\, C\,\,\, BA\,\,\, B\,\,\, A\,\,\, I]\\[3mm]
~~~+{\rm diag}[\Delta, 0,0,0,0,0,0,0]\\[3mm]
\equiv\Sigma+{\rm diag}[\Delta, 0,0,0,0,0,0,0],
\ea
$$
where $\Delta=E_1-A^{\dag}B^{\dag}C^{\dag}CBA$ (diag$[\sigma_1,
\sigma_2,\cdots]$ denotes a matrix with diagonal blocks $\sigma_1,
\sigma_2,\cdots $).
$\Sigma$ is PPT and has $7N$ kernel vectors:
$$
\ba{ll}
|001\ra|f\ra-|111\ra CB|f\ra,&~~~
|010\ra|g\ra-|111\ra CA|g\ra,\\[3mm]
|011\ra|h\ra-|111\ra C|h\ra,&~~~|100\ra|p\ra-|111\ra BA|p\ra,\\[3mm]
|101\ra|q\ra-|111\ra B|q\ra,&~~~|110\ra|m\ra-|111\ra A|m\ra,\\[3mm]
|000\ra|k\ra-|111\ra CBA|k\ra,
\ea
$$
for arbitrary $|f\ra,~|g\ra,~\cdots, ~|k\ra\in{\cal C}_{\sc d}^N$.

Taking into account $\rho \ge 0$, we have
$\Delta \ge 0$. Moreover, since $r(\rho)=r(\Sigma)$, the ranges of the
related matrices satisfy the relation $R(\rho)=R(\Sigma)\supseteq R({\rm
diag}[\Delta,\ 0])$. Therefore the corresponding kernels fulfill
$K({\rm diag}[\Delta,\ 0])\supseteq K(\Sigma)$. For vectors
$|\phi_k\ra$ belong to the kernel $K(\Sigma)$, we can deduce $\la
\phi_k|{\rm diag}[\Delta,\ 0]|\phi_k\ra=0$. As $\Delta \ge 0$,
we obtain that $\Delta|k\ra=0$ for all $|k\ra$, and thus $\Delta=0$.
\hfill $\Box$

Using Lemma we can prove the following Theorem:

{\bf Theorem.}\ \ A PPT-state $\rho$ in ${\cal C}^2 \otimes {\cal
C}^2 \otimes {\cal C}^2 \otimes {\cal C}^N$ with $r(\rho)=N$ is separable
if there exists a product basis $|e_A,\ f_B,\ g_C\ra$ such that
$r(\la e_A,\ f_B,\ g_C|\rho |e_A,\ f_B,\ g_C\ra)=N$.

{\bf Proof. }\ \ According to the Lemma the PPT state $\rho$ can be written as
\[
\rho=
\left(
\begin{array}{c}
A^{\dag}B^{\dag}C^{\dag}\\B^{\dag}C^{\dag}\\A^{\dag}C^{\dag}\\
C^{\dag}\\A^{\dag}B^{\dag}\\B^{\dag}\\A^{\dag}\\1
\end{array}
\right)
\left(
\begin{array}{cccccccc}
CBA&CB&CA&C&BA&B&A&1
\end{array}
\right).
\]
Since all $A$, $A^\dag$, $B$, $B^\dag$, $C$ and $C^\dag$ commute,
they have common eigenvectors $|f_n\ra$.
Let $a_n$, $b_n$ and $c_n$ be the corresponding eigenvalues of
$A$, $B$ and $C$ respectively. We have
\[
\la f_n|\rho |f_n\ra=
\left(
\begin{array}{c}
a_n^*b_n^*c_n^*\\[2mm]b_n^*c_n^*\\[2mm]a_n^*c_n^*\\[2mm]c_n^*\\[2mm]a_n^*b_n^*\\[2mm]
b_n^*\\[2mm]a_n^*\\[2mm]1
\end{array}
\right)
\left(
\begin{array}{cccccccc}
c_nb_na_n&c_nb_n&c_na_n&c_n&b_na_n&b_n&a_n&1
\end{array}
\right)\hskip5cm
\]
\[
=\left[\left(\begin{array}{c}c_n^*\\1\end{array}\right)\otimes
\left(\begin{array}{c}b_n^*\\1\end{array}\right)\otimes
\left(\begin{array}{c}a_n^*\\1\end{array}\right)\right]
(c_n\,\,\, 1)\otimes(b_n\,\,\, 1)\otimes(a_n\,\,\, 1)
=|e_{\sc a},f_{\sc b},g_{\sc c}\ra\la e_{\sc a},f_{\sc b},g_{\sc c}|.
\]
We can thus write $\rho$ as
$$\rho=\sum_{n=1}^N|\psi_n\ra\la \psi_n|\otimes |\phi_n\ra\la \phi_n|\otimes
|\omega_n\ra\la \omega_n|\otimes |f_n\ra\la f_n|,
$$
where
$$
|\psi_n\ra=\left(\begin{array}{c}c_n^*\\ 1\end{array}\right),~~~
|\phi_n\ra=\left(\begin{array}{c}b_n^*\\1\end{array}\right),~~~
|\omega_n\ra=\left(\begin{array}{c}a_n^*\\1\end{array}\right).
$$
Because the local transformations are reversible, we can now
apply the inverse transformations and obtain a decomposition of the initial
state $\rho$ in a sum of projectors onto product vectors. This
proves the separability of $\rho$. \hfill $\Box$

We have derived a canonical form of PPT states in
${\cal C}^2 \otimes {\cal C}^2\otimes {\cal C}^2 \otimes {\cal
C}^N$ with rank $N$. A necessary separability criterion is also obtained
from the representation of the canonical form. The results can be generalized
to multipartite quantum systems with more sub-systems like
${\cal C}^2 \otimes {\cal C}^2 \otimes \cdots \otimes {\cal C}^N$,
or with higher dimensions.

\end{document}